\documentclass[twocolumn,showpacs,preprintnumbers,amsmath,amssymb]{revtex4}
\usepackage{graphicx}

\def\lsim{\mathrel{\rlap{\lower4pt\hbox{\hskip1pt$\sim$}}
    \raise1pt\hbox{$<$}}}                
\def\gsim{\mathrel{\rlap{\lower4pt\hbox{\hskip1pt$\sim$}}
    \raise1pt\hbox{$>$}}}                

\begin{document}
\title{Scaling of load in communications networks}
\author{Onuttom Narayan$^1$ and Iraj Saniee$^2$}
\affiliation{$^1$ Department of Physics, University of California, Santa Cruz, CA 95064}
\affiliation{$^2$ Mathematics of Networks Department, Bell Laboratories, Alcatel-Lucent, 600 Mountain Avenue, Murray Hill, NJ 07974}
\date{\today}
\begin{abstract}
We show that the load at each node in a preferential attachment network
scales as a power of the degree of the node. For a network whose
degree distribution is $p(k)\sim k^{-\gamma},$ we show that the load is
$l(k)\sim k^\eta$ with $\eta = \gamma - 1,$ implying that the probability
distribution for the load is $p(l)\sim 1/l^2$ independent of $\gamma.$
The results are obtained through scaling arguments supported by finite
size scaling studies. They contradict earlier claims, but are in
agreement with the exact solution for the special case of tree graphs.
Results are also presented
for real communications networks at the IP layer, using the latest available
data. Our analysis of the data shows relatively poor power-law degree 
distributions as compared to the scaling of the load versus degree. 
This emphasizes the importance
of the load in network analysis.
\end{abstract}
\maketitle

A variety of problems in fields ranging from the social sciences
to biology to engineering deal with networks, for which a unified
understanding has been sought~\cite{bareview,dorog,newman}. One
of the models commonly used is the preferential attachment
(PA) model due to Barabasi and Albert~\cite{barabasi} and its
generalizations~\cite{russian,russian2,redner}. This model generates
scale-free networks, in which the probability for a node to have a degree
$k$ scales as $p(k)\sim k^{-\gamma}.$ Depending on the parameters of the
model, $\gamma$ can be varied continuously over the range $2 < \gamma
\leq 3.$ It has been argued that this model is appropriate to describe
communications networks such as the Internet.

Among the various properties of PA networks that have been studied
is the distribution of load (with uniform demand) at different nodes of the network. This
is defined by assuming that one unit of traffic flows between each
pair of nodes in the network along the shortest path connecting
them\cite{footbc}.  (If there are multiple shortest paths between a pair of nodes,
the traffic between them is divided equally among all the shortest
paths\cite{foot1}.) In this setting, the amount of traffic flowing through
a node is its load. Based on numerical simulations~\cite{goh}, it was claimed that the
probability distribution for the load scales as $p(l)\sim 1/l^\delta$
with $\delta = 2.2.$ Subsequently, data for networks in various different
fields were presented, and it was argued that~\cite{goh2} there are two
universality classes with $\delta = 2$ and $\delta = 2.2.$ These claims
were disputed~\cite{comment} on the basis further numerical simulations,
which seemed to indicate that $\delta$ varies continuously with $\gamma$
and is therefore not universal. For the special case of PA networks that
are trees, it was argued~\cite{szabo} and then proved~\cite{riordan}
that $\delta = 2.$

In this paper, we show that the average load at nodes of degree $k$
scales as $l(k)\sim k^\eta$ with $\eta = \gamma - 1$ for the PA model,
regardless of $\gamma.$ (As $k$ is increased for fixed $N,$ finite
size effects are seen.)  If we assume that the distribution of load
for fixed $k$ and $N$ does not have an anomalously large width, this
implies that the exponent $\delta$ is universal, but is equal to 2,
contradicting the earlier claims~\cite{goh,goh2,comment}, and extending
the exact result for PA trees.  We also extend the analytical proof of
Ref.~\cite{riordan} for tree graphs to show directly that $\eta = 2,$
supporting the assumption that the distribution of $l$ for fixed $k$ is
not anomalous.  Our results are obtained by simple scaling arguments that
are reinforced by finite size scaling studies. The deviations from this
universal result that are observed~\cite{goh,comment} are due to finite
size scaling and subleading corrections to the asymptotic scaling form.

We also show results for load analyses on networks drawn from a
recent database~\cite{rocketfuel} of connectivity of communications networks
at the IP layer. The data are collected with new measurement techniques, and 
find many more routers and links than earlier studies~\cite{f3}. The results demonstrate that the scaling of the load $l(k)\sim k^\eta$ is much
clearer than that of the ---more commonly studied--- degree distribution.

In the generalized PA model, a network grows one node at a time.
Each node is born with $m$ undirected edges which are attached to
preexisting nodes. The probability of attachment to a preexisting node
of degree $k$ is proportional to $k + k_0.$ Thus $k_0$ and $m$ are the
parameters of the model, with $k_0 < -m.$ For an infinite network, it
can be shown that the probability of a randomly chosen
node having a degree $k$ is to
\begin{equation}
p_k\propto k^{-\gamma}\qquad \gamma = 3 + k_0/m
\label{degdist}
\end{equation}
for large $k.$ For such a network, we assume ---as verified later
through numerical simulations--- that the average load $l_N(k)$ at all
the nodes of degree $k$ in a network of $N$ nodes has the scaling form
\begin{equation}
l_N(k) = N k^\eta\hat l (k/N^\mu)
\label{scaling1}
\end{equation}
where $\hat l(x)\rightarrow 1$ as $x\rightarrow 0$ and $\hat
l(x)\rightarrow 0$ as $x\rightarrow\infty.$ The prefactor of $N$ is
reasonable, since most of the load at nodes near the periphery of the
network, for which $k\sim O(1)$, is due to traffic that starts or ends
there, and is therefore $O(N).$ The exponent $\eta$ can be found by
noting that $\sum_k [N p_k] l_N(k)$ is the total traffic flowing in the
network. Since $N^2$ units of traffic are generated in the graph, and the
average geodesic length is $\sim \ln N,$ the sum should scale as $\sim
N^2 \ln N$ for large $N.$ From Eqs.(\ref{degdist}) and (\ref{scaling1}),
this implies that
\begin{equation}
\eta = \gamma - 1.
\label{eta}
\end{equation}
To find the exponent $\mu,$ we note that for a network of $N$ nodes,
the maximum degree $k_{max}$ that is achieved can be estimated by
requiring $(1 - \sum_{k_{max}}^\infty p_k)^N$ to be $\sim O(1).$ From
Eq.(\ref{degdist}), for large $k_{max}$ this is equivalent to $\exp[ -
A N/k_{max}^{\gamma - 1}]\sim O(1)$ with some constant $A,$ from which
$k_{max}\sim N^{1/(\gamma - 1)}.$ If we assume that all characteristic
$k$'s scale with $N$ in the same way, we obtain
\begin{equation}
\mu = 1/(\gamma - 1).
\label{mu}
\end{equation}
For $k << N^\mu,$ Eq.(\ref{scaling1}) implies that $l_N(k)\sim N k^\eta.$ 
When combined with Eq.(\ref{degdist}), we have $p(l) dl \sim dl/l^\delta$
with
\begin{equation}
\delta = \frac{\gamma - 1}{\eta} + 1 = 2
\end{equation}
where we have used Eq.(\ref{eta}). 

Although these results are plausible, they are based on assumptions, most
notably the scaling hypothesis of Eq.(\ref{scaling1}) itself.  To check
these assumptions, we turn to finite size scaling numerical simulations.
Networks with $N$ ranging from 500 to 8000 or 16000 were generated for
different values of $m$ and $k_0.$ We considered the cases $k_0 = 0$ with
$m=1,2,4,6$ and $m=6$ with $k_0 = 1,2,3,4,5.$ For each choice of $k_0,m$
and $N,$ 10000 graphs were generated, with traffic flowing as described
above. $l_N(k)$ was calculated by averaging over all the nodes of degree
$k$ in the 100 graphs. For the plots, it is more convenient to use the
form $l_N(k) = N^2 \tilde l(k/N^\mu)$ instead of Eq.(\ref{scaling1}),
which one obtains if one uses Eqs.(\ref{mu}) and (\ref{eta}).

\begin{figure}
\begin{center}
\includegraphics[width=\columnwidth]{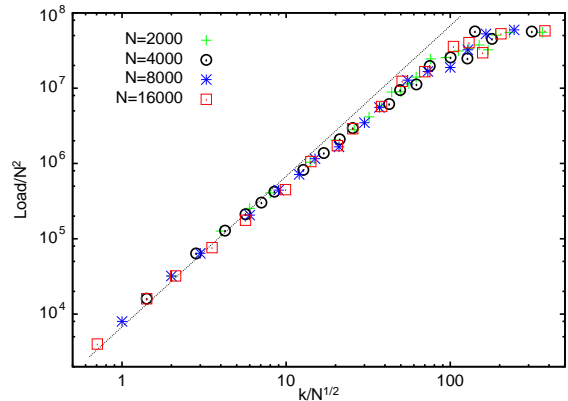}
\caption{Average traffic at nodes of degree $k$ as a function of $k$ for
the PA model with $m=1, k_0 = 0.$ (Here and in the subsequent figures,
$N/8000$ is used to scale both axes instead of $N.$) A scaling collapse
with the exponents from Eqs.(\ref{eta}) and (\ref{mu}) works
reasonably well. However, a straight line with the predicted slope of 2.0 is 
shown and only fits the curve --- if at all --- for small $k.$ 
Similar deviations are seen for $m=2,4,6$ with $k_0 = 0.$ }
\label{fig:offset0}
\end{center}
\end{figure}
Figure~\ref{fig:offset0} shows the results for $m=1$ and $k_0 = 0,$ for
which $\mu = 1/2$ from Eq.(\ref{mu}). The scaling collapse is reasonable,
but the best fit straight line is in fact $l(k)\sim k^1.8,$ consistent
with earlier results~\cite{szabo}. In view of the analytical results for
PA tree graphs, this discrepancy must be attributed to finite size scaling
effects which flatten the curve for large $k$ and presumably reduce the
apparent value of $\eta.$ The same discrepancy is seen for $k_0 = 0$
with $m=2,4,6$; it is reasonable to attribute it to the same cause.

\begin{figure}
\begin{center}
\includegraphics[width=\columnwidth]{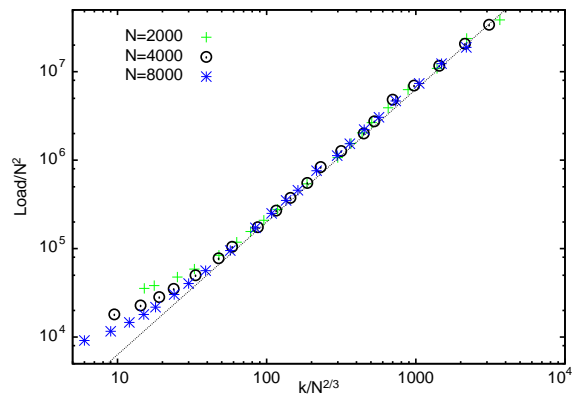}
\caption{Scaling plot of the average traffic at nodes of degree $k$
as a function of $k$ for the PA model with $m = 6,k_0 = 3.$ The scaling
collapse is very good, as is the fit to $\sim k^{3/2}$ in the scaling 
regime.}
\label{fig:offset3}
\end{center}
\end{figure}
\begin{figure}
\begin{center}
\includegraphics[width=\columnwidth]{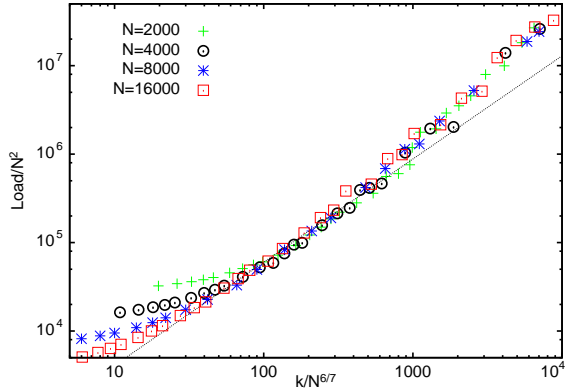}
\caption{A plot similar to Figure~\ref{fig:offset3} but with $k_0 = 5.$ For
$k\sim O(1),$ the individual curves pull away from the scaling form. For 
$k \sim O(N^{6/7}),$ finite size effects cause the curves to bend upwards. 
Between these two regimes, the slope is consistent with $\sim k^{7/6}$ as 
predicted.}
\label{fig:offset5}
\end{center}
\end{figure}

Figure~\ref{fig:offset3} shows a similar scaling plot for $m=6$ and
$k_0 = 3,$ for which $\mu = 2/3.$ The scaling collapse and the fit
to $l(k)\sim k^1.5$ are both very good in this case.  Figure~\ref{fig:offset5}
is a similar plot for $m=6$ and $k_0 = 5.$ The scaling collapse is
again very good, but finite size corrections for large $k$ now {\it
increase\/} the slope of the curve. Thus the fit to the predicted form
of $\sim k^{7/6}$ only works when $k << N^{6/7}$ but $k >> O(1).$ (
Unless the scaling hypothesis
breaks down, Eq.(\ref{eta}) follows from the $\ln N$ factor in the total
load. Therefore, we do not try a scaling plot with adjustable exponents..) We
have also made similar plots for $m=6$ and $k_0 = 1,2$ and 4. For $k_0 <
3,$ finite size corrections reduce the apparent $\eta$ for large $k,$
while for $k_0 > 3,$ they increase the apparent $\eta$ for large $k.$
This is consistent with Ref.~\cite{comment}.

For the tree graphs generated by the PA model with $m=1,k_0 = 0,$
the result $l(k)\sim k^2$ follows from $p(k)\sim 1/k^3$ and $p(l)\sim
1/l^2$ if we assume that $l(k)$ scales as a power of $k$ and that the
distribution of $l$ for fixed $k$ is not anomalously broad. Although
these are reasonable assumptions, it is not difficult to prove $l(k)\sim
k^2$ directly.  The probability that the node created at time $\tau$ will
be attached to a preexisting node of degree $k$ is equal to $k/(2\tau -
2).$ Therefore the probability that a node created at time $\tau$ will
have exactly $k$ nodes subsequently attached to it is
\begin{equation}
p_{k+1,N}(\tau) = 
\sum_{\tau < \tau_1 <\ldots \tau_k}^{N} 
P(\tau + 1, \tau_1 - 1, 1) \frac{1}{2(\tau_1 - 1)} 
P(\tau_1 + 1, \tau_2 - 1, 2)\frac{2}{2(\tau_2 - 1)} \ldots 
P(\tau_{k-1} + 1, \tau_k - 1, k) \frac{k}{2\tau_k - 2} P(\tau_k + 1, N, k + 1)
\label{probk}
\end{equation}
where
\begin{equation}
P(\tau, \tau^\prime, m) = 
\Bigg(1 - \frac{m}{2\tau - 2}\Bigg)\Bigg(1 - \frac{m}{2\tau}\Bigg)\ldots 
\Bigg(1 - \frac{m}{2\tau^\prime - 2}\Bigg). 
\end{equation}
Replacing $P(\tau,\tau^\prime,m)$ as the exponential of an
integral instead of a sum in the approximation $\tau >> 1,$ we have
$P(\tau,\tau^\prime, m) = (\tau/\tau^\prime)^{m/2}.$ With this,
Eq.(\ref{probk}) simplifies to
\begin{equation}
p_{k+1,N}(\tau) 
\approx \sum_{\tau_1 = \tau}^N\ldots\sum_{\tau_k = \tau}^N 
\sqrt\frac{\tau}{N}\frac{1}{2\sqrt{\tau_1 N}}\ldots \frac{1}{2\sqrt{\tau_k N}}
\label{pktau}
\end{equation}
where we have used the symmetry of the $\tau_i$'s to eliminate
the restriction $\tau_1 < \tau_2\ldots < \tau_k.$ If the sums
are replaced with integrals, $p_{k+1,N}(\tau) = \sqrt{\tau/N} [1 -
\sqrt{\tau/N}]^k$\cite{kr} and $p_k = (1/N)\sum_{\tau} p_{k,N}(\tau)\sim
1/[k (k + 1) (k + 2)]$ for large $N.$

If $n_1, n_2,\ldots n_k$ are the sizes of the $k$ subtrees descending from
a node of degree $k+1,$ one can show~\cite{riordan} that for large $N$
the load at the node is proportional to $N\sum n_i.$ If $n_i(t)$ is the
size of the $i$'th subtree at time $t\geq \tau_i,$ with $n_i(N) = n_i,$
the probability that $n_i(t + 1) = n_i(t) + 1$ is $(2 n_i - 1)/(2t),$
with the initial condition $n_i(t) = 1.$ at time $t.$ Averaging over
randomness for fixed $\tau_i,$ the solution is $\langle n_i(t)\rangle =
(t/\tau_i + 1)/2.$ With the symmetrization of the previous paragraph,
\begin{equation}
l_N(k+1;\tau) \propto N\langle \sum n_i\rangle = N k \langle n_i\rangle 
= \frac{kN}{1-x}\int_x^1 \frac{1 + x_i^2}{x_i^2} dx_i
\end{equation}
where $\tau_i = N x_i^2, \tau = N x^2$ and we have replaced sums with
integrals. This yields
\begin{equation}
l_N(k+1) 
\propto N k^3\int_0^1  x (1 - x)^k [ 2 k (1 + x) + x ] dx
\propto N k^2
\label{last}
\end{equation}
where we have only kept the terms that are relevant for $N>>
k >> 1.$ The terms dropped with the $k >> 1$ approximation are
corrections to asymptotic scaling and cause the imperfect collapse in
Figure~\ref{fig:offset0} .

\begin{figure}[htb]
\begin{center}
\includegraphics[width=\columnwidth]{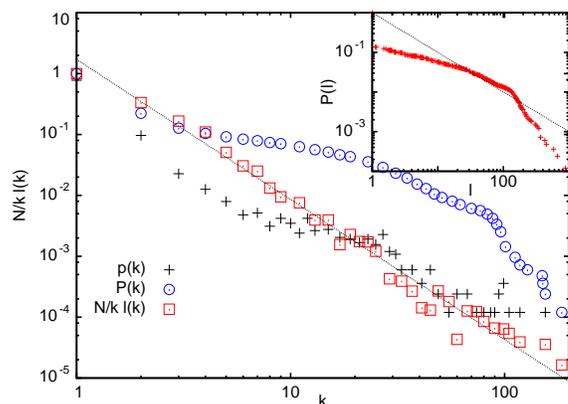}
\caption{Log log plot of $N/(k l(k))$ as a function of $k$ for the
Sprintlink network~\cite{rocketfuel}.  The dashed line is visually
adjusted for best fit, and has a slope of $-2.3$ corresponding to
$l(k)\sim k^\eta$ with $\eta= 1.3.$ The plot also shows the degree
distribution $p(k)$ and the cumulative distribution $P(k) = \sum_k^\infty
p(k_1).$ The inset shows the cumulative load distribution $P(l) =
\int_l^\infty p(l_1) dl_1$ and a straight line with slope -1.}
\label{fig:sprintlink}
\end{center}
\end{figure}
\begin{figure}
\begin{center}
\includegraphics[width=\columnwidth]{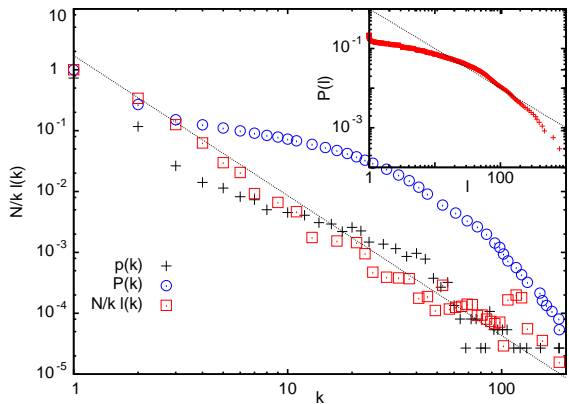}
\caption{Plot similar to Figure~\ref{fig:sprintlink} but with results from all
ten networks in the database combined.}
\label{fig:rocketfuel}
\end{center}
\end{figure}

We now compare with network data from the Rocketfuel
database~\cite{rocketfuel}, which is the most recent, comprehensive and 
publicly available collection of measurements of the connectivity between 
nodes of communications networks at the IP layer.
There are ten networks with 121 to 10214
routing nodes (from here on referred to as "routers")
in the database. Since the data is insufficient to test scaling
functions (in our simulations we worked with 100 networks for each
$N,$ with $500 < N < \sim 10000$), we only consider how the traffic
at nodes scales with their degree without regard to any finite size
cutoff. Figure~\ref{fig:sprintlink} shows the results for one of the
larger networks in the database, the Sprintlink network with 8355 routers.
The load as a function of degree fits quite well to a power law with
a slope of 1.3. The degree distribution itself is much more irregular,
and it is difficult to say whether it is of the form $\sim k^{-\gamma}$
with $\gamma = \eta + 1.$
The cumulative degree distribution is also shown
in the figure; it is much smoother, but does not show clear power law
behavior. The inset shows the cumulative load distribution and a straight
line with the predicted slope of -1, which is equally unconvincing.
Figure~\ref{fig:rocketfuel} is a similar figure with all ten networks
in the database merged. The plots are straighter, and
one could perhaps argue for a narrow power law regime in the cumulative degree
distribution as is done in Ref.~\cite{rocketfuel} or a 
regime with slope $-1$ in the inset. However, the
load versus degree has a much closer linear behavior in the log-log plot, and 
is therefore a more natural demonstration of power-law scaling in 
communications networks than the more commonly studied
degree distribution.

In conclusion, we have shown that for a preferential attachment model
with degree distribution $p(k)\sim k^{-\gamma},$ the average traffic
as a function of node degree scales as $l(k) \sim k^{\gamma - 1}.$
This is equivalent to the statement that the probability distribution
for the load scales as $p(l)\sim 1/l^2$ regardless of $\gamma.$ Although
the numerical simulations and analytical calculations are for a specific
model, the result follows from the scaling assumption and the small-world
phenomenon and is therefore more robust.
The scaling $l(k)\sim k^\eta$ is also seen clearly in networks at the
IP layer, and is in fact much better than the degree distribution which
has attracted much more interest.

This work was supported by AFOSR grant FA9550-08-1-0064.


\begin{thebibliography}{}
\bibitem{footbc} This notion of load is sometimes referred to as
``betweenness centrality".
\bibitem{foot1} We do not distinguish between geodesics that are partly
overlapping and those that are completely distinct. Thus if there are
three geodesics between a pair of nodes, with the first two overlapping
for most of their lengths and the last one completely separate, the
traffic along each would be $1/3,$ resulting in greater traffic over
most of the overlapping geodesics.
\end{thebibliography}
\end{document}